\documentclass[11pt]{article}
\usepackage{mymoriond,epsfig}

\def\ra{\rightarrow}

\def\be{\begin{equation}}
\def\ee{\end{equation}}
\def\bea{\begin{eqnarray}}
\def\eea{\end{eqnarray}}

\def\eqs#1#2{Eqs. (\ref{#1}) and (\ref{#2})}
\def\tab#1{Table~\ref{#1}}
\def\sec#1{Section~\ref{#1}}
\def\fig#1{Figure~\ref{#1}}

\def\ts{\vrule height2.5ex depth0pt width0pt}

\newcommand{\PBS}[1]{\let\temp=\\#1\let\\=\temp}% latex companion p108
\def\gtrsim{\mathrel{\lower .7ex\hbox{$\buildrel\textstyle>\over\sim$}}}
\def\lesssim{\mathrel{\lower .7ex\hbox{$\buildrel\textstyle<\over\sim$}}}

\newcommand{\fbd}{f_{B_d}}
\newcommand{\fbs}{f_{B_s}}

\newcommand{\bbd}{B_{B_d}}

\newcommand{\rgbbd}{\hat{B}_{B_d}}

\def\gtrsim{\mathrel{\lower .7ex\hbox{$\buildrel\textstyle>\over\sim$}}}
\def\lesssim{\mathrel{\lower .7ex\hbox{$\buildrel\textstyle<\over\sim$}}}

\let\errp\errparen

\makeatletter
\def\slash#1{{\mathpalette\c@ncel{#1}}} % TeXbook, bottom of p360
\makeatother
\def\l{\left}
\def\r{\right}
\def\la{\langle}
\def\ra{\rangle}
\def\ord#1{{\mathcal O}\l(#1\r)}

\def\beq{\begin{equation}}
\def\eeq{\end{equation}}
\def\bea{\begin{eqnarray}}
\def\eea{\end{eqnarray}}

\def\gev{\mbox{GeV}}
\def\mev{\mbox{MeV}}
\def\msbar{{\overline{\mathrm{MS}}}}

\makeatletter
\def\slash#1{{\mathpalette\c@ncel{#1}}} % TeXbook, bottom of p360
\makeatother

\begin{document}
\begin{flushright}
CERN-TH/99-140\\
CPT-99/PE.3817
\end{flushright}
%\vspace*{4cm}
\title{{\Large$\Delta m_d$}, {\Large $\Delta m_s/\Delta m_d$} {\large AND} 
{\Large $\epsilon_K$} {\large IN QUENCHED
QCD}
\footnote{Invited talk given at the {\it XXXIVth Rencontres de Moriond: 
Electroweak Interactions and Unified Theories, Les Arcs, France, 
13-20 March 1999}}}

\author{LAURENT LELLOUCH
\footnote{On leave from Centre de Physique Th\'eorique, Case
        907, CNRS Luminy, F-13288 Marseille Cedex 9, France.}}

\address{TH Division, CERN, CH-1211 Geneva 23, Switzerland}

\maketitle\abstracts{ I present quenched, lattice QCD calculations of
the hadronic matrix elements relevant for $B^0_{d(s)}-\bar
B^0_{d(s)}$ and $K^0-\bar K^0$ mixing and briefly review the status
of lattice predictions.}

%\newpage
\section{Introduction}

Neutral meson mixing is a rich source of information on the Standard
Model (SM). For instance, the frequencies with which $B_d$ and $B_s$
mesons oscillate into their anti-particles yield constraints on the
Cabibbo-Kobayashi-Maskawa (CKM) matrix element $V_{td}$ which
determines the most poorly known side of the unitarity triangle.
$K^0-\bar K^0$ mixing, on the other hand, through its measured
contribution to indirect CP violation in $K\to\pi\pi$ decays, provides
a constraint on the triangle's summit. These constraints require
quantification of the non-perturbative QCD dynamics which modify the
simple, underlying quark processes. The uncertainties in this
quantification must be reduced to allow for as stringent a test of the
SM as possible with the triangle's angles soon to be measured at the
$B$-Factories, HERA, the Tevatron and the LHC. Lattice QCD provides a
first principle tool which can help achieve this goal. In what
follows, I present results of lattice calculations performed with
C.J.~David Lin and the UKQCD Collaboration
%~\footnote{Preliminary
%results and more details can also be found in
%\cite{Lellouch:1998xk,Lellouch:1998sg}} 
as well as a summary of
lattice predictions~\footnote{For other recent lattice reviews,
please see \cite{reviews}.}.

\section{$B^0_{d,(s)}-\bar B^0_{d,(s)}$ mixing}

$B^0_q$ and $\bar B^0_q$ ($q{=}d,s$) are not eigenstates of the weak
hamiltonian and can therefore oscillate into one another with a
frequency given by the mass difference, $\Delta m_q$, of the
eigenstates of the full SM hamiltonian. In the SM, the dominant
contribution to this mass difference is given by~\cite{Buras:1997fb}
\beq
\Delta m_q \simeq \frac{G_F^2}{8\pi^2}\, M_W^2\, |V_{tq}
V_{tb}^*|^2\, S_0\l(x_t\r)
\eta_BC_B(\mu)\, 
\frac{|\langle \bar B_q|O^{\Delta B=2}_q(\mu)|B_q\rangle|}
{2M_{B_q}}
\ , \quad O^{\Delta B=2}_q=
[\bar
b\gamma^\mu(1-\gamma^5) q][\bar b\gamma_\mu(1-\gamma_5) q]\ .
\label{eq:dmq}
\eeq
$x_q\equiv (m_q^2/M_W^2)$ and $\eta_B$, $S_0$ and $C_B$ are
short-distance quantities, calculated perturbatively. Thus, $|V_{tq}|$
can be determined from a measurement of $\Delta m_q$ once the
non-perturbative matrix element $\langle \bar B_q|O^{\Delta
  B=2}_q(\mu)|B_q\rangle$ is quantified~\footnote{Assuming
  three-generation unitarity and present day constraints on CKM
  parameters, $|V_{tb}|=1$ to high accuracy.}. This is where the
lattice enters.

\begin{figure}
\centerline{
\epsfxsize=10cm\epsfysize=5cm\epsffile{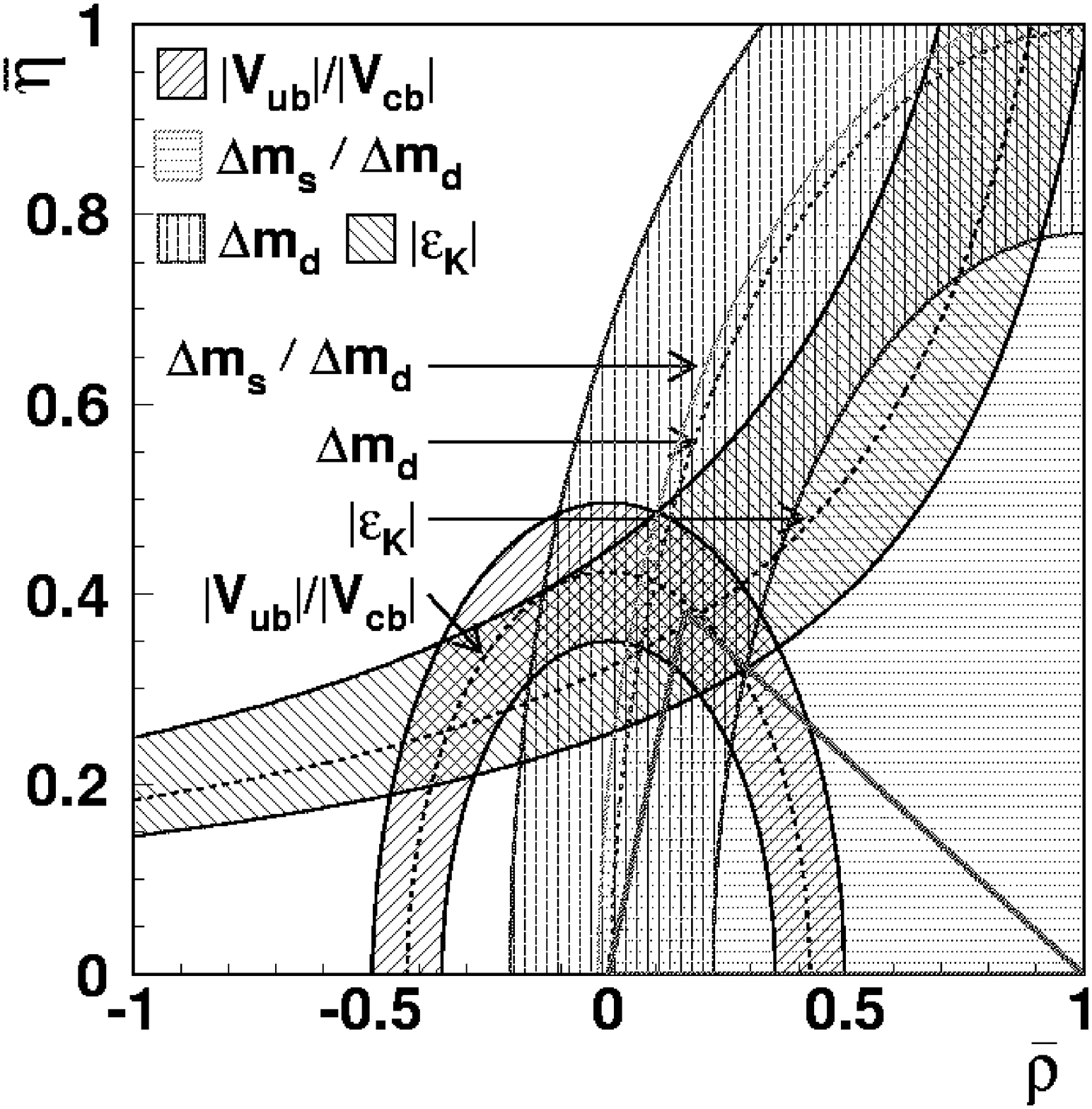}}
\caption{Constraints on 
  the Wolfenstein parameters $\bar\rho$ and $\bar\eta$ from $\Delta
  m_d$, $\Delta m_s/\Delta m_d$ and
  $\epsilon_K$~\protect\cite{Mele:1999bf} (for illustration only).
\label{fig:triangle}}
\end{figure}

While the measurement of $\Delta m_d$ provides a direct determination
of $|V_{td}|$, one may also consider the ratio
\beq
\frac{\Delta m_s}{\Delta m_d}=\l|\frac{V_{ts}}{V_{td}}\r|^2\
\frac{M_{B_d}}{M_{B_s}}\l|\frac{\langle \bar B_s|O^{\Delta B=2}_s|B_s\rangle}
{\langle \bar B_d|O^{\Delta B=2}_d|B_d\rangle}\r|
\equiv\l|\frac{V_{ts}}{V_{td}}\r|^2\,\frac{M_{B_d}}{M_{B_s}}\,r_{sd}
\equiv \l|\frac{V_{ts}}{V_{td}}\r|^2
\frac{M_{B_s}}{M_{B_d}}\,\xi^2
\ .
\label{eq:rsddef}
\eeq
This ratio gives another possible constraint on $|V_{td}|$ since
with three generations, $|V_{ts}|\simeq |V_{cb}|$. It further has the
advantage that many common factors and uncertainties in the
evaluation of the matrix elements cancel. Measuring $\Delta m_s$,
however, remains an experimental challenge, as the neutral $B_s$
mesons oscillate rapidly: $\Delta m_s\ge 12.4\mbox{ ps}^{-1}$ at 95\%
CL versus $\Delta m_d= 0.477(17)\mbox{ ps}^{-1}$ for $B_d$
mesons~\cite{lepboscichep98}. Nevertheless, even this lower bound on $\Delta
m_s$ provides a significant constraint on $|V_{td}|$ as can be seen
in \fig{fig:triangle}.

Traditionally, the matrix element $\langle\bar B_q|O^{\Delta
  B=2}_q(\mu)|B_q\rangle$ is normalized by its vacuum saturation
value:
\beq
B_{B_q}(\mu)\equiv \frac{\langle\bar B_q|O^{\Delta B=2}_q(\mu)|B_q\rangle}
{\langle\bar B_q|O^{\Delta B=2}_q|B_q\rangle_{VSA}} =
\frac{\langle\bar B_q|O^{\Delta B=2}_q(\mu)|B_q\rangle}
{(8/3) M_{B_q}^2 f_{B_q}^2}
\ .\eeq
While one can actually determine the matrix element itself on the
lattice, $B$-parameters are obtained from ratios of correlation
functions in which many statistical and systematic uncertainties
are expected to cancel. Furthermore, the matrix element has mass dimension four
and therefore suffers very strongly from the uncertainty associated
with the determination of the lattice cutoff which is of order 10\% in
present day quenched calculations. As we shall also see later, it
is advantageous to get the matrix element from an independent
determination of $B_{B_q}$ and $f_{B_q}$ and the experimental value of
$M_{B_q}$.

\subsection{A parte on decay constants}
\label{sec:decaycsts}

Because the leptonic decay constants of $B$ mesons are required, I
briefly digress to comment on their values. Many lattice groups have
calculated these constants over the years. A recent compilation can
be found in~\cite{lhc99} where the following summary numbers,
which include uncertainties due to quenching, are given:
\beq
\fbd  =  175 \pm 35\,\mev\ ,\quad
\fbs =  200 \pm 35\,\mev\quad\mbox{and}\quad
\fbs/\fbd  =  1.14 \pm 0.08 \ ,
\label{eq:deccst}
\eeq
in a normalization where $f_\pi=131\,\mev$. While
the effects of quenching in $(f_{B_s}/f_{B_d})$ appear to be small
in simulations~\cite{Bernard:1998xi}, Quenched $\chi$PT (Q$\chi$PT)
indicates that they could be significant~\cite{Sharpe:1996qp}.

\subsection{$\Delta B=2$ matrix elements and $B$-parameters}
\label{sec:dbeq2}

In \tab{tab:bparam}, I present our results for $B_{B_d}$ and
$B_{B_s}/B_{B_d}$ along with a compilation of results obtained by other
groups who use, as we do, ``relativistic'' heavy quarks, as opposed to
NRQCD or static quarks. (Details of our calculations can be found in
the Appendix.)
\begin{table}
\caption{\label{tab:bparam} Results for $B$-meson $B$-parameters obtained 
with ``relativistic'' heavy quarks. 
$\beta$ is the coupling at which the calculations were performed.
$\beta=\infty$ corresponds to results extrapolated to the continuum
limit. $\mu$ is the matching scale used. The numbers in italics are
derived from the published results. Running is performed at two-loops
using the procedure of J. Flynn {\it et al.}~\protect\cite{reviews} which
assumes $m_b=5\,\gev$. $\rgbbd^{\mathrm{nlo}}$ is the
RG-invariant $B$-parameter at NLO.}
\begin{center}
\begin{tabular}{@{}rcclllll@{}}
\hline
\ts & action & $\beta$ & $\mu[\gev]$ & $\bbd(\mu)$ & $\bbd(m_b)$
 & $\rgbbd^{\mathrm{nlo}}$ & $B_{B_s}/\bbd$ \\[0.4ex]
\hline
\ts UKQCD~\cite{Lellouch:1998xk} & 
MFI SW & 6.2 & 2.6 & 0.95(3) & 0.91(3) & 1.45(5) & 0.99(3)\\
(preliminary) & & 6.0 & 2.0 & 0.94(5) & 0.89(4) & 1.41(7) & 1.03(3)\\
BBS98~\cite{Bernard:1998dg} & Wilson & $\infty$ & 2 & 1.02(13) & \sl0.96(12) &
  \sl1.53(19) & $\sim$ \sl1\\
JLQCD96~\cite{Aoki:1995jq} & Wilson & 6.3 & & & 0.840(60) & \sl1.34(10) & $\sim$ \sl1.05\\
& & 6.1 & & & 0.895(47) & \sl1.42(7)  & $\sim$ \sl0.99\\
BS96~\cite{Soni:1996qq} & Wilson & $\infty$ & 2 & 0.96(6)(4) & \sl0.90(6)(4) &
 \sl1.44(9)(6) & 1.01(4)\\
ELC92~\cite{Abada:1992mt} & Wilson & 6.4 & 3.7 & 0.86(5) & \sl0.84(5) &
  \sl1.34(8) \\
BDHS88~\cite{Bernard:1988dy} & Wislon & 6.1 & 2 & 1.01(15) & 
\sl0.95(14) & \sl1.51(22)
 \\[0.4ex]
\hline
\end{tabular}
\end{center}
\end{table}

Quenching errors for these $B$-parameters and $B$-parameter ratios
have been studied with the help of Q$\chi$PT \cite{Sharpe:1996qp} and
have been found to be small.  Combining this information with the
results of \tab{tab:bparam}, I give the following estimates:
\beq
\rgbbd^{\mathrm{nlo}}=1.4(1)\quad\quad\mbox{and}\quad\quad
\frac{B_{B_s}}{B_{B_d}}=1.00(3)
\ .
\label{eq:bparavg}
\eeq

In order to use these results to extract $|V_{td}|$ from a measurement of
$\Delta m_d$, we need to combine them with a determination of $f_{B_d}$. Using
the estimate given in \sec{sec:decaycsts}, I quote:
\beq
f_{B_d}\sqrt{\rgbbd^{\mathrm{nlo}}}=207(42)\,\mev
\ .\eeq
This prediction can be compared to the value obtained from an
overconstrained, unitarity-triangle fit with
$f_{B_d}\sqrt{\rgbbd^{\mathrm{nlo}}}$ left as a fit
parameter~\cite{Parodi:1999nr}:
$f_{B_d}\sqrt{\rgbbd^{\mathrm{nlo}}}=223(13)\,\mev$.  This fit
incorporates lattice predictions for $r_{sd}$ and $B_K$ consistent
with the ones given below. Agreement is excellent, indicating a
general consistency of the SM and lattice calculations.  The central
value and error bars, of course, reflect the choices made by the
authors for the various inputs they use.

\subsection{$SU(3)$ breaking in $\frac{\Delta m_s}{\Delta m_d}$}

There are at least two possible ways of obtaining $r_{sd}$
from the lattice:
\begin{itemize}
\item[a)] taking the product
\beq
r_{sd}^{(a)}\equiv \l(\frac{M_{B_s}}{M_{B_d}}\r)^2\l(\frac{f_{B_s}}
{f_{B_d}}\r)^2
\l(\frac{B_{B_s}}{B_{B_d}}\r)
\ ,
\eeq
with $\l(f_{B_s}/f_{B_d}\r)$ and $\l(B_{B_s}/B_{B_d}\r)$ 
determined on the lattice
and $\l(M_{B_s}/M_{B_d}\r)$ measured experimentally;

\item[b)] from a direct determination of the ratio
\beq
r_{sd}^{(b)}\equiv \l|\frac{\langle \bar B_s|O^{\Delta B=2}_s|B_s\rangle}
{\langle \bar B_d|O^{\Delta B=2}_d|B_d\rangle}\r|
\ .\eeq
\end{itemize}
Our results for $r_{sd}$, together with the results of other groups
who use ``relativistic'' heavy quarks are summarized in \tab{tab:rsd}.
\begin{table}
\caption{\label{tab:rsd} 
Results for $r_{sd}$ as obtained using methods a) and b) with
``relativistic'' heavy quarks.}
\begin{center}
\begin{tabular}{rcccc}
\hline
& action & $\beta$ & $r_{sd}^{(a)}$ & $r_{sd}^{(b)}$ \\[0.4ex]
\hline
UKQCD~\cite{Lellouch:1998xk} & MFI SW  & 6.2 & 1.37(13) & 1.70(28) \\
(preliminary) & & 6.0 & 1.38(7) & 1.52(19) \\
BBS98~\cite{Bernard:1998dg} & Wilson & $\infty$ & 1.42(5)\errp{28}{15} &
1.76(10)\errp{57}{42}\\[0.4ex]
\hline
\multicolumn{3}{c}{w/ results of \protect\eqs{eq:deccst}{eq:bparavg}} 
& 1.34(19) &\\[0.4ex]
\hline
\end{tabular}
\end{center}
\end{table}
Comparison of $r_{sd}^{(a)}$ at our two values of the lattice spacing
($\beta=6.2$ and 6.0) suggests that discretization errors are small.
Furthermore, we find that $r_{sd}^{(a)}$ and $r_{sd}^{(b)}$ are
compatible, though the latter is less accurate and less reliable: its
heavy-quark and light-quark-mass dependences are stronger and
the corresponding extrapolations are less well controlled.

On the basis of these results and the comments on quenching in
Sections \ref{sec:decaycsts} and \ref{sec:dbeq2}, I quote as summary
values:
\beq
r_{sd} = 1.4(2)\quad\quad\mbox{or}\quad\quad \xi\equiv \sqrt{r_{sd}}
\l(\frac{M_{B_d}}{M_{B_s}}\r)=1.16(8)
\ .\eeq

\section{$K^0-\bar K^0$ mixing}

$K^0-\bar K^0$ mixing induces indirect CP violation in $K\to\pi\pi$
decays, quantified by the parameter $\epsilon_K$~\cite{Buras:1997fb}:
\beq
\epsilon_K\, e^{-i\frac{\pi}{4}}
%\equiv \frac{{\mathcal A}(K_L\to(\pi\pi)_{I=0})}
%{{\mathcal A}(K_S\to(\pi\pi)_{I=0})}
\simeq C_\epsilon\, C_K(\mu)B_K(\mu)A^2\lambda^{10}\bar\eta\l[(1-\bar\rho) A^2
\eta_2S_0(x_t)+P_0(x_t,x_c,\ldots)\r]
= (2.280\pm 0.013)\times 10^{-3}
\ ,
\eeq
with
\beq
\la\bar K^0|O_{\Delta S=2}(\mu)|K^0\ra=
\frac{8}{3}M_K^2\,f_K^2\times B_K(\mu)
\quad\quad\mathrm{and}\quad\quad
O_{\Delta S=2}=[\bar s\gamma_\mu(1-\gamma^5) d]
[\bar s\gamma^\mu(1-\gamma^5) d]
\ .\eeq
This in turn leads to a hyperbolic constraint on the summit
$(\bar\rho,\bar\eta)$ of the unitarity triangle, once the
$B$-parameter $B_K$ is determined (see \fig{fig:triangle}).  Here,
$C_\epsilon$ is obtained from well mesured quantities, $A$ and
$\lambda$ are Wolfenstein parameters and $\eta_2$, $C_K$, $S_0$ and
$P_0$ incorporate perturbative, short-distance physics ($P_0$ also
contains CKM factors). We calculate $B_K$ on the same lattices as the
$\Delta B=2$ matrix elements.

\subsection{Chiral subtractions}

Even though the basic ingredients, such as the operator mixing
alluded to in the Appendix, are very similar to those used to
calculate the $\Delta B=2$ matrix elements, the physics here is very
different, as it is governed by chiral symmetry. In the continuum,
$O_{\Delta S=2}$ is in the $(27,1)$ representation of $SU(3)_L\times
SU(3)_R$. On the lattice, however, the explicit breaking of chiral
symmetry implies the following chiral expansion:
\beq
\la\bar K^0(\vec{q})|O_{\Delta S=2}|
K^0(\vec{p})\ra_{lat}
= \alpha_K+\beta_K\,M_K^2+\gamma_K\,(p\cdot q)+\cdots
\ ,\ee
where $\alpha_K$ and $\beta_K$ are pure lattice artefacts, while
$\gamma_K(p\cdot q)$ and higher-order terms contain the physical
contributions.  In our calculation, where we match onto the continuum
at one loop, the artefacts $\alpha_K$ and $\beta_K$ are proportional
to $\alpha_s^2$ and $a\alpha_s$.  The problem is that even though
these factors are small, the physical contributions are chirally
suppressed compared to $\alpha_K$.

To quantify and subtract the unphysical contributions, we study the
chiral behavior of the $\Delta S=2$ matrix element as a function of
$M_K^2$ and $p\cdot q$.  At $\beta=6.2$ we find that artefacts such as
$\alpha_K$ and $\beta_K$ are small and consistent with zero for all
matching scales in the range $1/a\to\pi/a$. We have checked that our
results are robust to procedure by normalizing the $\Delta S=2$ matrix
element in a variety of ways and using different mass and recoil
variables for the chiral expansion. The determination of $B_K$ from
the corresponding physical expansion terms should thus be reliable. At
$\beta=6.0$, the lattice artefacts are around 2 standard deviations
away from zero and the results are less robust to procedure. Our
findings, together with results obtained with less improved actions,
suggest that discretization errors represent an important part of the
traditionally observed residual chiral violations.

\subsection{Results for $B_K$}

We take our $\beta=6.2$ result as our best estimate. We run it to
$2\,\gev$ at two-loops with $n_f=0$ in the $\msbar$-NDR scheme (small
running). Our results, together with those of other groups are
summarized in \tab{tab:bkresults}.
\begin{table}
\caption{\label{tab:bkresults} Results for $B_K^{(NDR)}(2\,\gev)$.
The results in the second half of the table were obtained with
discretizations of the quark action which maintain a partial or
full chiral symmetry, obviating the need for chiral subtractions.}
\begin{center}
\begin{tabular}{cccc}
\hline
Ref. & action & $\beta$ & $B_K^{(NDR)}(2\,\gev)$\\
\hline
UKQCD~\cite{Lellouch:1998sg} (preliminary) 
& MFI SW & 6.2 & $0.72^{+8}_{-6}$\\
APE98~\cite{Allton:1998sm} & SW & 6.0, 6.2 & 0.68(21)$^{(a)}$\\
GBS97~\cite{Gupta:1997yt} & Wilson & 6.0 & 0.74(4)(5)\\
JLQCD99~\cite{Aoki:1999gw} & Wilson & $\infty$ & 0.69(7)\\
\hline
JLQCD98~\cite{Aoki:1998nr} &  Staggered & $\infty$ & 0.628(42)\\
KGS98~\cite{Kilcup:1997ye} & Staggered & $\infty$ & 0.62(2)(2)\\
BS97~\cite{Blum:1997mz} & Domain Wall & $\infty$ &0.628(47)$^{(b)}$\\
\hline
\multicolumn{3}{l}{{\small $^{(a)}$ matched to NDR; 
$^{(b)}$ matched at tree level}}
\end{tabular}
\end{center}
\end{table}
On the basis of $\chi$PT and preliminary unquenched
results~\cite{Kilcup:1997hp}, Sharpe estimates that $SU(3)$-breaking
corrections~\footnote{Calculations are performed with 
  degenerate or 
nearly degenerate $s$ and $d$ quarks.} and unquenching may lead to an
$\ord{10\%}$ increase in $B_K$ and ascribes an $\ord{15\%}$ error to
$B_K$ to account for the uncertainties in this
estimate~\cite{reviews}. Bijnens {\it et al.}~\cite{Bijnens:1995br}
reach similar conclusions. I choose to include these effects as a
contribution to the error but not to the central value.

On the basis of these conclusions and the results given in
\tab{tab:bkresults}, I quote:
\beq
B_K^{(NDR)}(2\,\gev)=0.65(10)\quad\mbox{or}\quad\hat B_K^{\mathrm{nlo}}=
0.89(14)\ ,
\eeq
where $\hat B_K^{\mathrm{nlo}}$ is the two-loop RG-invariant
$B$-parameter obtained from $B_K^{(NDR)}(2\,\gev)$ with $n_f=3$ and
$\alpha_s(2\,\gev)=0.3$. Again, this result compares very favorably to
the SM fit of~\cite{Parodi:1999nr}, but this time with $\hat
B_K^{\mathrm{nlo}}$ as a fit parameter instead of
$f_{B_d}\sqrt{\rgbbd^{\mathrm{nlo}}}$: $\hat
B_K^{\mathrm{nlo}}=0.87^{+0.34}_{-0.20}$.

\section{Conclusions}

The lattice provides a means for calculating $\Delta B=2$ and $\Delta
S=2$ matrix elements from first principles. A reliable determination
of these matrix elements will be crucial for testing the SM with the
forthcoming experiments on CP violating $B$ decays. Moreover, the
hadronic matrix elements which appear in supersymmetric extensions of
the SM can also be considered~\cite{Allton:1998sm}.  In the next few
years, more and more unquenched calculations will be performed,
enabling a better quantification of quenching effects and eventually
yielding fully unquenched results.

\section*{Acknowledgments}
I wish to thank my collaborator David Lin as well as Jonathan Flynn and
Guido Martinelli for many useful discussions on the topics presented
here. Support from EPSRC and PPARC under grants GR/K41663 and
GR/L29927, and from the EEC through TMR network EEC-CT98-00169 is
acknowledged.

\section*{Appendix: details of the calculations}

We describe quarks with a mean-field-improved, Sheikholeslami-Wohlert
(MFI SW) action. Compared to the standard Wilson action, the leading
discretization errors are formally reduced by a factor of order
$\alpha_s(a)$, and the mean-field-improvement may give additional
numerical suppression. We perform calculations at two values of the
cutoff, corresponding to couplings $\beta\equiv
3/(2\pi\alpha_s(a))=6.0$ (coarser lattice) and 6.2 (finer lattice).
This enables us to quantify discretization errors. The parameters of
the simulations are summarized in \tab{tab:simparam}.
\begin{table}
\caption{Parameters of our calculations. ``$\#$ cfs'' is the number
of gauge-field configurations on which the various matrix elements
are computed (i.e. our statistics). $c_{SW}$ is the
mean-field-improved coefficient of the SW term. $a^{-1}(m_\rho)$ is
the inverse lattice spacing as determined from a calculation of the
$\rho$-meson mass. \label{tab:simparam}}
%\vspace{0.4cm}
\begin{center}
\begin{tabular}{ccccc}
\hline
$\beta$ & lattice size & $\#$ cfs & $c_{SW}$ &
$a^{-1}(m_\rho)$\\
\hline
6.2 & $24^3\times 48$ & 188 & 1.442 & $2.57(8)\ \gev$\\
6.0 & $16^3\times 48$ & 498 & 1.479 & $1.96(5)\ \gev$\\
\hline
\end{tabular}
\end{center}
\end{table}
Note that our simulations have high statistics.  Unfortunately,
because of the very high numerical cost of including the feedback of quarks on
the gauge-fields, both calculations are performed in the quenched
approximation.

Because a physical pion would feel the boundaries of the box in which
we work and because the algorithms we use slow down rapidly for
lighter quarks, we are restricted to work with quarks with masses on
the order of $m_s/2$ or more. Thus, to obtain results at the
physical values of the $u,\,d$ and $s$ masses, we perform all
calculations for three values of the light-quark mass roughly in the range
$m_s/2\to m_s$ and extrapolate or interpolate to the physical mass values.

Furthermore, the graininess of our lattice forbids us from working with
quarks whose masses are much larger than $m_c$. Thus, we perform all
calculations for five values of the heavy-quark mass around that of
the charm and extrapolate to the physical $b$-quark mass.

Finally, because Wilson fermions break chiral symmetry explicitely,
the left-left operator, $O_{\Delta F=2}$, mixes with four-quark
operators of different chirality. We subtract these wrong chirality
contributions and match to the $\msbar$-NDR scheme at one loop. To
estimate the systematic error associated with our procedure, we vary
the matching scale in the range $1/a\to\pi/a$. For the $B$-parameters
discussed here, dependence on this scale is very small.

\end{document}